\documentclass[preprint,onecolumn,amsmath,amssymb,floatfix,nofootinbib,secnumarabic]{revtex4}
\usepackage{graphics}      % standard graphics specifications
\usepackage{graphicx}      % alternative graphics specifications
\usepackage{longtable}     % helps with long table options
\usepackage{url}           % for on-line citations
\usepackage{bm}            % special 'bold-math' package
\usepackage[dvips]{epsfig,color}
\usepackage{dcolumn}

\def \c{\hat{c}}
\def \a{\hat{a}}
\def \d{\hat{d}}
\def\S{\boldsymbol{S}}
\def \beq{\begin{equation}}
\def \eeq{\end{equation}}
\def \beqarr{\begin{eqnarray}}
\def \eeqarr{\end{eqnarray}}
\def \bspt{\begin{split}}
\def \espt{\end{split}}
\def \bef{\begin{figure}}
\def \enf{\end{figure}}
\def \tr{\text{tr}}

\newcommand{\abs}[1]{\lvert#1\rvert}

\newcommand{\braket}[2]{\mbox{$\langle #1 | #2 \rangle$}}
\newcommand{\meanvalue}[3]{\mbox{$\langle #1 | #2 | #3 \rangle$}}
\newcommand{\proj}[2]{\mbox{$|#1\rangle \!\langle #2 |$}}
\newcommand{\ev}[1]{\mbox{$\langle #1 \rangle$}}
\newcommand{\bra}[1]{\mbox{$\langle #1 |$}}
\newcommand{\ket}[1]{\mbox{$| #1 \rangle$}}
\newcommand{\nmod}[1]{\mbox{${^*_*} #1 {^*_*}$}}

\begin{document}

\title{Entanglement Entropy in Many-Fermion System}

\author{Wenxin Ding}
\affiliation{NHMFL and Department of Physics, Florida State
University, Tallahassee, Florida 32306, USA}

\begin{abstract}
In this four-part prospectus, we first give a brief introduction to
the motivation for studying entanglement entropy and some recent
development. Then follows a summary of our recent work about
entanglement entropy in states with traditional long-range
order. After that we demonstrate calculation of entanglement entropy in
both one-dimensional spin-less fermionic systems as well as bosonic
systems via different approaches, and connect them using
one-dimensional bosonization. In the last part, we briefly sketch
the idea of bosonization in high-dimensions, and discuss the
possibility and advantage of approaching the scaling behavior of
entanglement entropy of fermions in arbitrary dimensions via bosonization.
\end{abstract}

\date{August 20, 2008}

\maketitle

\tableofcontents

\newpage

%%%%%%%%%%%%%%%%%%%%%%%%%%%%%%%%%%%%%%%
%                                 %%%%%
%         Introduction            %%%%%
%                                 %%%%%
%%%%%%%%%%%%%%%%%%%%%%%%%%%%%%%%%%%%%%%

\section{Introduction}
Entanglement is the hallmark as well as most counter-intuitive feature
of quantum mechanics. Its contradiction with one of the most important
concepts in classic physics - locality - has caused skepticism and
controversy\cite{EPR} ever since quantum mechanics was formulated. It
was only after the genuine insight of Dr. John Bell\cite{bell_87},
where based on several general assumptions he derived a set of
inequalities for two physical observables that any local theory
should obey,that  the difference between truly quantum entanglement
and classic or local correlation can be tested and distinguished
experimentally. However, the experiments only become viable after
decades\cite{PhysRevLett.28.938} since these inequalities were
formulated. The overwhelming majority of experiments done nowadays
support quantum entanglement\cite{peres93}. Even though there are
still critics pointing out that these Bell test experiments are not
problem-free, the existence of such experimental problems which are often
referred to as 'loopholes' may affect the validity of the experimental
findings, they are still demonstrating that quantum entanglement is
physical reality.

Entanglement has attracted more attention due to the development of
quantum information and quantum computation science\cite{nielsen2000}
where entanglement is considered major resource of quantum
information.

Among various ways of quantifying entanglement, bipartite block
entanglement entropy has emerged as a concept of central importance,
not only in quantum information science, but recently in other branches of
physics as well. In condensed matter/many-body physics the
entanglement entropy has been increasingly used as a very useful and
in some cases indispensable way to characterize phases and phase
transitions, especially in strongly correlated phases
\cite{amico-2007}.

This bipartite block entanglement entropy is defined as the following. For
any given density operator of a pure state $\rho$, we can divide the system
into two parts, then we partially trace out either part from the
division:

\beq
\rho_1 = tr_2(\rho),\ \rho_2 = tr_1(\rho).
\eeq

And the entanglement entropy is defined as von Neumann entropy of the
reduced density matrix:

\beq\label{entanglemententropy}
E = E_1 = E_2 = - tr(\rho_1 \ln \rho_1) = - tr(\rho_2 \ln \rho_2)
\eeq

In this context the most important result is
perhaps the so-called area law\cite{PhysRevD.34.373, area-law}, which
states that in the thermodynamic limit, the entropy should be
proportional to the area of the boundary that divides the system into
two blocks. There are a very few important examples\cite{amico-2007,
  wolf:010404, gioev:100503} in which the area law is violated, most of which
involves quantum criticality\cite{wilczek94, PhysRevLett.90.227902,
  calabrese-2004-0406, PhysRevLett.93.260602, santachiara-2006,
  feiguin:160409, bonesteel:140405}; the specific manner with
which the area law is violated is tied to certain universal
properties of the phase or critical point. In some other cases,
important information about the phase can be revealed by studying the
leading {\it correction} to the area law\cite{kitaev:110404,
 levin:110405, fradkin:050404}; for example this is the case for
topologically ordered phases\cite{kitaev:110404, levin:110405}.

It is to our particular interest to study the entanglement entropy of
many-fermion systems which are most encountered in condensed matter
physics. There have been numerous works on this topic, mostly
numerical. %( references need to be added here)
There are also some analytic results in 1D spin-less fermions, based on
conformal field theory argument\cite{wilczek94, calabrese-2004-0406}
or fermionic wave
functions\cite{its-2005-38, Jin2004}, giving entanglement entropy $~
\frac{1}{3}\log L$ where $L$ is the subsystem size. For higher
dimensions, Wolf\cite{wolf:010404} establishes a relation between structure
of the Fermi sea and the scaling of the entropy for a finite nonzero
Fermi surface:

\beq
S \sim L^{d-1}\log L.
\eeq

while Gioev and Klich\cite{gioev:100503} provide an explicit geometric formula for
the entropy of free fermions in any dimension $d$ as $L \rightarrow
\infty$ by making a connection with Widom's conjecture\cite{widom}:

\beq
S \sim \frac{L^{d-1}\log L}{(2 \pi)^{d-1}}\frac{1}{12} \int_{\partial
  \Omega} \int_{\partial \Gamma} \abs{n_x \cdot n_p} dS_x dS_p,
\eeq

%%%%%%%%%%%%%%%%%%%%%%%%%%%%%%%%%%%%%%%%%%%%%%%%%%%%%%%%%%%%%
%%%%%%%%%%%%%%%%%%%%%%%%%%%%%%%%%%%%%%%%%%%%%%%%%%%%%%%%%%%%%

\subsection{General Form of Reduced Density Matrix}
In this section, we will consider a general formalism for obtaining
the reduced density matrix in Fock space.

Consider a general pure state $\ket{\Psi_0}$, and we want to divide it into
two parts in some orthonormal basis. Let the basis be $\ket{B}$, and
the basis of the two divisions are denoted as $\ket{e}$, $\ket{s}$
respectively. Here we use $e$ indicating the environment which we want
to trace out and $s$ denoting the subsystem of interest. Then in
general we can write:

\beq\label{Psi0}
\ket{\Psi_0} = \sum_{B} c_B \ket{B} = \sum_{e,s} c_{e,s}\ket{e}
\otimes \ket{s}.
\eeq

So the density matrix of the whole system is written as:

\beq
\rho_0 = \proj{\Psi_0}{\Psi_0} = \left( \sum_{e_1,s_1} c_{e_1,s_1} \ket{e_1}
\otimes \ket{s_1} \right) \left( \sum_{e_2,s_2} c_{e_2,s_2}^* \bra{e_2} \otimes \bra{s_2}\right).
\eeq

In order to easily trace out the environment part, we collect the
states with the same environment parts, which yields:

\beq
\rho_0 = \left( \sum_{e_1} (\sum_{s_1} c_{e_1, s_1} \ket{s_1}) \otimes
\ket{e_1} \right) \left( \sum_{e_2} (\sum_{s_2} c_{e_2, s_2}^*
\bra{s_2}) \otimes \bra{e_2} \right).
\eeq

Then we move on to the calculation of the reduced density matrix by
tracing out the environment:

\beq
\rho_s = tr_e(\rho_0) = \sum_{e_0} \left( \bra{e_0} ( \sum_{e_1}
(\sum_{s_1} c_{e_1, s_1} \ket{s_1}) \otimes
\ket{e_1} ) ( \sum_{e_2} (\sum_{s_2} c_{e_2, s_2}^*
\bra{s_2}) \otimes \bra{e_2} ) \ket{e_0} \right).
\eeq

Since {$\ket{e}$} is chosen to be orthonormal, so $\braket{e_0}{e_1} =
\delta_{e_0,e_1}$. the above equation is reduced to:

\begin{equation}\label{reduceddensitymatrix}
\begin{split}
 \rho_s & = tr_e(\rho_0) = \sum_{e_0} \left( (\sum_{s_1} c_{e_0, s_1}
\ket{s_1}) (\sum_{s_2} c_{e_0, s_2}^*
\bra{s_2}) \right) \\
&= \sum_{s_1,s_2} (\sum_{e_0} c_{e_0,s_1} c_{e_0,s_2}^*) \proj{s_1}{s_2}.\\
\end{split}
\end{equation}

So from Eq.\ref{reduceddensitymatrix} we can get that the matrix element of the reduced
density matrix of subsystem $s$ is:

\beq
(\rho_s)_{(s_1,s_2)} = \meanvalue{s_1}{\rho_s}{s_2} = \sum_{e}
c_{e,s_1} c_{e, s_2}^*
\eeq

According to Eq.\ref{Psi0}, we can write:
\beq
(\rho_s)_{(s_1,s_2)} = \sum_{e} c_{e,s_1} c_{e,s_2}^* =
\meanvalue{\Psi_0}{\hat{a}^\dag_{s_2} \hat{a}_{s_1}}{\Psi_0},
\eeq

where $\hat{a}_{s}$, which we shall call 'state annihilation
operator', is defined within the Hilbert space of the subsystem as
$\hat{a}_{s_0}\ket{s} = \delta_{s_0,s} \ket{0}$, $\ket{0}$ being the vacuum
state of the subsystem. The specific form of $\hat{a}_{s_0}$ need be
addressed in specific systems.

Now we can write the reduced density matrix in a compact form:

\beq
\rho_s = \sum_{s_1,s_2} \meanvalue{\Psi_0}{\hat{a}^\dag_{s_2}
  \hat{a}_{s_1}}{\Psi_0} \hat{a}^\dag_{s_1} \hat{a}_{s_2}
\eeq

This formalism is general and works for any complete basis of the
Fock space of concern.

As long as we are only concerning about the density matrix of the
subsystem, we can replace our 'state annihilation operators' by any
complete set of operators that span the complete Hilbert space of the
subsystem.

%%%%%%%%%%%%%%%%%%%%%%%%%%%%%%%%%%%%%%%
%                                 %%%%%
% EE in States Long-Range Order   %%%%%
%                                 %%%%%
%%%%%%%%%%%%%%%%%%%%%%%%%%%%%%%%%%%%%%%

\section{Entanglement Entropy in States with Traditional Long-Range Magnetic Order}

A lot of works have been done on entanglement entropy in various
aspects, however, there have been relatively few studies of the behavior
of entanglement entropy in states with traditional long-range
order\cite{vidal05,vidal06,vidal07}. This is perhaps because of the
expectation that ordered states can be well described by mean-field
theory, and in mean-field theory the states reduce to simple product
states that have no entanglement. In particular in the limit of
perfect long-range order the mean-field theory becomes ``exact", and
the entanglement entropy should vanish. In this section, we will
summarize our recent work\cite{wenxin} on those states and show
that this is {\em not} the case, and interesting entanglement exists
in states with perfect long-range order. We will study two exactly
solvable spin-1/2 models: (i) An unfrustrated antiferromagnet with
infinite range (or constant) antiferromagnetic (AFM) interaction
between spins in opposite sub-lattices, and ferromagnetic (FM)
interaction between spins in the same sub-lattice; (ii) An ordinary
spin-1/2 ferromagnet with arbitrary FM interaction among the
spins\cite{popkov:012301}\footnote{The entanglement property of this
  model has bee previously studied in Ref.\cite{popkov:012301}. Here
  we introduce a different definition of the entanglement entropy that
  properly takes into account the ground state degeneracy, and can be
  calculated much more straightforwardly using symmetry
  consideration. See Sec. 2.3.}. While the ground states have perfect
long-range order for both models, we show that they both have non-zero
entanglement entropy that grow logarithmically with the size of the
subsystem.

%%%%%%%%%%%%%%%%%%%%%%%%%%%%%%%%%%%%%%%%%%%%%%%%%%%%%%%%%%%%%%%%%%%%%%%%%%%%%%%
%%%%%%%%%%%%%%%%%%%%%%%%%%%%%%%%%%%%%%%%%%%%%%%%%%%%%%%%%%%%%%%%%%%%%%%%%%%%%%%

\subsection{Antiferromagnetic Spin Model and the Ground State}
\begin{figure}
  \includegraphics[width=16cm]{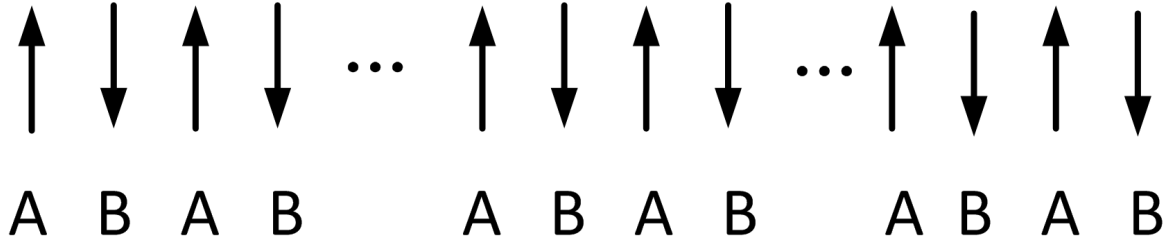}\\
  \caption{Two-sub-lattice model: two sub-lattices labeled A and B
  interpenetrating each other.}\label{Fig.lattice}
\end{figure}

We consider a lattice model composed of two sub-lattices
interpenetrating each other as in Fig. \ref{Fig.lattice}, with
interaction of infinite range, i.e., every spin interacts with all
the other spins in the system, with interaction strength
independent of the distance between the spins. Within each
sub-lattice, the interaction is ferromagnetic, and between the
sub-lattices the interaction is antiferromagnetic; as a result
there is no frustration. The Hamiltonian is written as,
%eq 1
\begin{equation}
\label{Ham}
H=-J_A\sum_{i,j\in A}\S_i \cdot \S_j-J_B\sum_{i,j\in B}\S_i
\cdot \S_j+J_0\sum_{i\in A,j\in B}\S_i \cdot \S_j,
\end{equation}
with $J_A, J_B, J_0 > 0$. The ground state of (\ref{Ham}) can be
solved in the following manner\cite{yusuf}. Introduction the following
notations $$\S_A =
\sum_{i\in A} \S_i,\ \ \ \S_B=\sum_{i\in B}\S_i,\ \ \
\S=\S_A+\S_B,$$ then the Hamiltonian can be written as,
%eq 2
\begin{equation}
\begin{split}
H &= -J_A\S_A^2-J_B\S_B^2+J_0\S_A\cdot\S_B\\
  &= -J_A\S_A^2-J_B\S_B^2 + \frac{J_0}{2}(\S^2-\S_A^2-\S_B^2).\\
\end{split}
\end{equation}

Then it is not hard to see that the ground state should be:
%eq 4
\begin{equation}
|S_AS_B;Sm\rangle = \bigg|\frac{N_A}{2},\frac{N_B}{2};
 \bigg|\frac{N_A}{2} - \frac{N_B}{2}\bigg|, m \bigg\rangle.
\end{equation}

%fig.1
%end fig.1

For simplicity we only consider the simplest case with $N_A = N_B =
N$, thus the total system size is $2N$. Then the ground state is
reduced to an antiferromagnetic ground state which has zero total
spin,
\beq \ket{S_AS_B;Sm}=\ket{\frac{N}{2}\frac{N}{2};00}.
\label{singletneel}
\eeq
This ground state has perfect Neel order, as manifested by the
spin-spin correlation function,
\begin{equation}
\begin{split}
&\langle \S_i\cdot\S_j\rangle ={1\over 4},\hskip 0.4 cm i, j \in A
  \hskip 0.2 cm {\rm or} \hskip 0.2 cm i, j \in B;\\
&\langle \S_i\cdot\S_j\rangle =-{1\over 4}-{1\over 4N}\rightarrow
  -{1\over 4}, \hskip 0.2 cm i\in A \hskip 0.2 cm {\rm and} \hskip 0.2
  cm j \in B.
\end{split}
\end{equation}

%%%%%%%%%%%%%%%%%%%%%%%%%%%%%%%%%%%%%%%%%%%%%%%%%%%%%%%%%%%%%%%%%%%%%%%%%%%%%
%%%%%%%%%%%%%%%%%%%%%%%%%%%%%%%%%%%%%%%%%%%%%%%%%%%%%%%%%%%%%%%%%%%%%%%%%%%%%

\subsection{Reduced density matrix and Entanglement Entropy}

We divide the system spatially into two subsystems which are labeled 1
and 2 respectively and study the ground state entanglement entropy $E$
between these two subsystems.

Following is a brief summary about how to solve for the explicit form
of the reduced density matrix.

First, we further decompose the system into four parts, $\S_{A_1},\
\S_{A_2},\ \S_{B_1},\ \S_{B_2}$, with
\beq
\begin{cases}
  \S_{A_1} = \sum_{i\in A \wedge i \in 1} \S_i,\ \ \S_{A_2} = \sum_{i\in
  A \wedge i \in 2} \S_i;\\
\S_{B_1} = \sum_{i\in B \wedge i \in 1} \S_i, \ \ \S_{B_2} =
  \sum_{i\in B \wedge i \in 2} \S_i.
\end{cases}
\eeq
Therefore, these operators satisfy the following relations,
\begin{equation}
  \begin{cases}
    \S_A = \S_{A_1}+\S_{A_2},\ \ \ \S_B = \S_{B_1}+\S_{B_2};\\
    \S_1 = \S_{A_1}+\S_{B_1},\ \ \ \S_2 = \S_{A_2}+\S_{B_2}.\\
  \end{cases}
\end{equation}
%%%%%%%%%%%%
%newly added
%%%%%%%%%%%%%%%%%%%%%%%%%%%%%%%%%%%%%%%%%%%%%%%%%%%%%%%%%%%%%%%%%%
Here we note that, as discussed in the previous section, the spin
state within each sub-lattice is ferromagnetic. This means that not
only must the total spin quantum numbers of $\S_A$ and $\S_B$ take their
maximum values, but the total spin quantum numbers of $\S_{A_1},\
\S_{A_2},\ \S_{B_1}$ and $\S_{B_2}$ must also take their maximum
values. More importantly, these values are thus fixed, which enables
us to treat the operators $\S_{A_1},\ \S_{A_2},\ \S_{B_1}$ and
$\S_{B_2}$ as four single spins, and in what follows we shall denote
these operators by their corresponding spin quantum numbers. The
problem is then that we are given a four spin state in which the spins
$S_{A_1}$ and $S_{A_2}$ are combined into a state with total spin
$S_A$ and the spins $S_{B_1}$ and $S_{B_2}$ are combined in a state
with total spin $S_B$, and then these two states are combined into a
total singlet (resulting in the ground state of our long-range AFM
model), and we must express this state in a basis in which the spins
$S_{A_1}$ and $S_{A_2}$ are combined into a state with total spin
$S_1$ and $S_{B_1}$ and $S_{B_2}$ are combined into a state with total
spin $S_2$.  This change of basis involves the familiar LS-jj coupling
scheme. At this point, to obtain the reduced density matrix we only
have to re-express the density matrix in bases of $\bra{S_1m_1S_2m_2;Sm}$ by
means of LSjj coupling, then trace out either $\bra{S_1 m_1}$ or $\bra{S_2m_2}$.

After some algebra we arrive at:

\beq \rho_1 = tr_{(2)}(\rho) = \sum_{S_1 m_1} |\lambda_{S_1
m_1}|^2 \ket{S_1 m_1} \bra{S_1 m_1}. \eeq

where $\lambda_{S_1, m_1}$ is given by
\begin{equation}\label{eqn:lambda}
\begin{split}
 \lambda_{S_{1},m_{1}} &=
 \bigg[\frac{(N+1)(N-N_1)!N_1!}{(N-N_1-S_{1})!(N-N_1+S_{1}+1)!}\\
& \times \frac{(N-N_1)!N_1!}{(N_1-S_{1})!(N_1+S_{1}+1)!}\bigg]^\frac{1}{2}.
\end{split}
\end{equation}

The bipartite entanglement entropy between subsystems 1 and 2 is
then given by,
\begin{equation}\label{eqn:entropy}
E= E_1 = -\sum_{S_1,m_1} \abs{\lambda_{S_1 m_1}}^2\ln(\abs{\lambda_{S_1 m_1}}^2).
\end{equation}

Here we note that, although $\lambda_{S_1m_1}$ is written with an
explicit $m_1$ dependence, the actual expression is independent of
$m_1$. As a result, we can eliminate the summation over $m_1$ from
(\ref{eqn:entropy}) by multiplying by a factor of $2S_1+1$.  The
final expression for the entanglement entropy is then,
\begin{equation}\label{eqn:entropy}
E= E_1 = -\sum_{S_1}(2S_1 + 1) \abs{\lambda_{S_1}}^2\ln(\abs{\lambda_{S_1}}^2).
\end{equation}

Next we shall present the asymptotic behavior of this bipartite
entanglement entropy in two limiting cases:
\begin{itemize}
\item{$N_1 = N_2 = N$, this case gives the saturated entropy at fixed $N$ since
  intuitively $E$ should increase with the subsystem size.}
\item{ $1 \ll N_1 \ll N$, in this limit we are considering
  system's entanglement with its (much larger) environment, and generically
we should be able
  to find that the entropy should be independent of the total system
  size as $N \rightarrow \infty$, which is indeed what we find.}
\end{itemize}

By extracting the asymptotic behavior of the eigenvalues of the
reduced density matrix $\lambda_{S_1}$, turning the summation into
integral and forcing the normalization condition of those eigenvalues,
the entanglement entropy can be analytically calculated:
\begin{equation}
  \begin{split}
    E_1 \simeq & - \int_0^\infty
    A(2S_1+1)e^{-S^2_{1}(\frac{1}{N-N_1}+\frac{1}{N_1})} \\ & \times \ln \left(Ae^{-S^2_{1}(\frac{1}{N-N_1}+\frac{1}{N_1})}\right) dS_1 \\
    \simeq & \ln\left(N_1 - \frac{N_1^2}{N} + \frac{1}{2}\sqrt{\frac{\pi (N-N_1)N_1}{N}}\right).
  \end{split}
\end{equation}

First let us consider the equal partition case, which presumably gives
the upper limit of the entanglement entropy for a given total system size.
For simplicity we set $N$ to be even (note that the total system size
is $2N$), thus $N_1 = \frac{N}{2}$, and
$\frac{1}{A}=\frac{1}{4}(N+\sqrt{N\pi})$. So the entanglement entropy becomes,
\begin{equation}
\begin{split}
  E &= \ln \left(\frac{N}{4} + \frac{1}{4}\sqrt{\pi N} \right)\\
& \simeq \ln N -\ln 4 \simeq \ln N - 1.38629.\\
\end{split}
\end{equation}

For the unequal partition case, $1 \ll N_1 \ll N$, we can expand the entropy as
follows,
\beq
\begin{split}
E &\simeq \ln \left[ N_1 \left(1 - \frac{N_1}{N} + \frac{1}{2}
\sqrt{\frac{\pi (1 -
    \frac{N_1}{N})}{N_1}}\right)\right]\\
& = \ln N_1 + \ln \left(1 - \frac{N_1}{N} + \frac{1}{2}
\sqrt{\frac{\pi (1 -
    \frac{N_1}{N})}{N_1}}\right)\\
& \simeq \ln N_1 - \frac{N_1}{N} +
O\left(\left(\frac{N_1}{N}\right)^2\right).
\end{split}
\eeq
From this expression we see that when the assumed condition is
satisfied the entropy indeed depends only on the subsystem size to
leading order.

\subsection{Ferromagnetic model and its entanglement entropy}

In this section we consider a ferromagnetic (FM) spin-1/2 model on
an arbitrary lattice with $N$ sites,
\beq H=-\sum_{i\ne
j}J_{ij}\S_i\cdot\S_j, \eeq
with $J_{ij} > 0$. The ground state is the fully magnetized state
$|SM\rangle$ with $S=N/2$ and $M=-S, -S+1, \cdots, S$, and is
clearly long-range ordered: $\langle\S_i\cdot\S_j\rangle=1/4$.
However there is a crucial difference between the FM ground state
and the AFM ground state studied earlier: the FM ground state has
a finite degeneracy, and thus the system exhibits a non-zero
entropy even at zero temperature, $E_{0} =\log(2S+1)=\log(N+1)$,
resulting from the density matrix of the entire system,
\beq
\rho =\frac{1}{2S+1}\sum_{M=-S}^S | SM \rangle \langle SM | =
\frac{1}{N + 1} \sum_{M=-S}^S | SM \rangle \langle SM |
\eeq

In this case the entanglement entropy between two subsystems (1 and 2)
is defined in the following manner. We first obtain reduced density
matrices for subsystems 1 and 2 by tracing out degrees of freedom in
2 and 1 from $\rho$:
\beq
\begin{split}
&\rho_1=tr_{(2)}\rho = \frac{1}{N+1} \sum_{M = -S}^{S} tr_{(2)} (| SM \rangle
\langle SM |) \\
&= \frac{1}{N+1} \sum_{M = -S}^{S} \rho_{1M},\\
&\rho_2= tr_{(1)}\rho = \frac{1}{N+1} \sum_{M = -S}^{S} tr_{(1)} (| SM
\rangle \langle SM |)\\
&= \frac{1}{N+1} \sum_{M = -S}^{S} \rho_{2M},
\end{split}
\eeq and calculate from them the entropy of the subsystems, $E_1$ and
$E_{2}$. The entanglement entropy is defined
as\cite{dreissig}\footnote{This definition is probably not unique. It
  is one half of the ``mutual information" introduced in
  Ref. \onlinecite{dreissig, Wolf2008}, and reduces to
  Eq. (\ref{entanglemententropy}) when $\rho$ is that of a pure
  state. The same definition was used by Castelnovo and Chamon [Claudio Castelnovo, and
  Claudio Chamon, Phys. Rev. B {\bf 76}, 174416 (2007)]. The name
  ``mutual information" may be first coined by Adami
  and Cerf [C. Adami and N.J. Cerf, Phys. Rev. A {\bf 56}, 3470
  (1997)] and Vedral, Plenio, Rippin and Knight [V. Vedral,
  M.B. Plenio, M.A. Rippin, and P.L. Knight, Phys. Rev. Lett. {\bf
  78}, 2275 (1997)], although Stratonovich [R. L. Stratonovich,
  Izv. Vyssh. Uchebn. Zaved., Radiofiz. {\bf 8}, 116 (1965);
  Probl. Inf. Transm. {\bf 2}, 35 (1966)] considered this quantity
  already in the mid-1960s.}
\beq E = (E_{1} + E_{2} - E_{0}) / 2. \eeq

For the present case $E_{1}$ and $E_{2}$ can be easily obtained
from the following observations. (i) Because the total spin is
fully magnetized, so are those in the subsystems: $S_1=N_1/2$ and
$S_2=N_2/2$. Thus this is a two-spin entanglement problem. (ii)
Because the total density matrix $\rho$ is proportional to the
identity matrix in the ground state subspace, it is invariant
under an arbitrary rotation in this subspace. (iii) As a result
the reduced density matrix $\rho_1$ is also invariant under
rotation in the subspace of subsystem 1 with $S_1=N_1/2$, and is
proportional to the identity matrix in this subspace. Thus \beq
\begin{split}
\rho_1 &= \frac{1}{N_1 + 1} \sum_{M_1=-\frac{N_1}{2}}^\frac{N_1}{2} |
\frac{N_1}{2}M_1 \rangle \langle \frac{N_1}{2}M_1 | \\
\end{split}
\eeq
and $E_{1}=\log(N_1+1)$ (in agreement with Ref.
\cite{popkov:012301}). Similarly $E_{2}=\log(N_2+1)$. Thus
\beq E =
[\log(N_1+1)+\log(N_2+1)-\log(N+1)]/2.
\eeq
We find in both the equal partition ($N_1=N_2=N/2$) and unequal
partition ($N_1\ll N_2=N-N_1$) limits, the entropy grows
logarithmically with subsystem size $N_1$,
\beq \lim_{N_1\rightarrow\infty}E \rightarrow {1\over 2}\log(N_1).
\eeq

%%%%%%%%%%%%%%%%%%%%%%%%%%%%%%%%%%%%%%%
%                                 %%%%%
% EE in 1D Fermions               %%%%%
%                                 %%%%%
%%%%%%%%%%%%%%%%%%%%%%%%%%%%%%%%%%%%%%%

\section{Entanglement Entropy of One Dimensional Spinless Free Fermions}

%%%%%%%%%%%%%%%%%%%%%%%%%%%%%%%%%%%%%%

\subsection{Fermionic Wave Function Approach}

%%%%%%%%%%%%%%%%%%%%%%%%%%%%%%%%%%%%%%

%%%%%%%%%%%%%%%%%%%%%%%%%%%%%%

\subsubsection{Reduced Density Matrix of Spinless Free Fermion on a 1D Lattice}

Now let us consider the application of this general formalism to the
case of free fermions on a one dimensional lattice of infinite
length. The subsystem is taken to N consecutive lattice sites. The
Hamiltonian is simply:

\beq\label{freeH}
\mathcal{H} = -t\sum_{n} (\hat{c}_{n+1}^\dag \hat{c}_{n} + \c_{n}^\dag
\c_{n+1}).
\eeq

The Hamiltonian is simply diagonalized by performing Fourier
transformation. let

\beqarr
\c_{n} &=& \frac{1}{\sqrt{2 \pi}}\int_{-\pi}^{\pi} dk e^{-ikn}\tilde{c}_k     \\
\tilde{c}_k &=& \frac{1}{\sqrt{2 \pi}} \sum_{n = -\infty}^{\infty} e^{ikn}\c_{n}.\\
\eeqarr

We get:

\beq
\mathcal{H} = -2t \int_{-\pi}^{\pi} dk \cos k \tilde{c}_k^\dag \tilde{c}_k.
\eeq

The ground state we will be interested in is to fill the vacuum state
up to the Fermi momentum:

\beq
\begin{split}
\ket{\Psi_0} &= \prod_{k = -k_f}^{k_f} \tilde{c}_k^\dag \ket{0} =
\left( \prod_{k} (\frac{1}{\sqrt{2 \pi}}\sum_{n} e^{-ikn} \c_n^\dag)
\right) \ket{0} \\
\end{split}
\eeq

But the mode of interest in our problem is the real space state, and
according to our general formalism, we shall try to find out our
'state annihilation operator'. For a finite lattice of length N, any
state can be written in such a general form:

\beq
\ket{\{n_i\}} = \prod_{i} (\c^\dag_{i})^{n_i} \ket{0} = \c_{\{n_i\}}^\dag \ket{0},
\eeq
where $i = 1,2,...,N$, $n_i = \{0, 1\}$, $\{n_i\}$ denoting a set of occupation
numbers.

However, since the ground state $\ket{\Psi_0}$ is not vacuum state for
$\c_i$, the 'state annihilation operators' can be spanned as $\a_s =
\prod_{i} \hat{O}_i$, with $\hat{O}_i \in \{\c_i,
\c^\dagger_i\}$. Noticing that Wick's theorem is valid in this system,
one can verify our assertion at the end of last section, and represent
$\rho_s$ as

\beq
\rho_s = tr_e (\rho_0) = \sum_{\{\hat{O}_i\}} \ev{(\prod_{i} \hat{O}_i)^\dagger}\prod_{i} \hat{O}_i,\
\text{with } \hat{O}_i \in \{\c_i,\c_i^\dagger,\c_i^\dagger \c_i,\c_i \c_i^\dagger\}
\eeq

Suppose we can find a unitary transformation: $\d_m = \sum_{n=1}^{L}
v_{mn} c_n$, so that $\ev{\d_i \d_j}=0,\ \meanvalue{\Psi_0}{\d_i^\dagger
  \d_j}{\Psi_0} = \delta_{ij} \meanvalue{\Psi_0}{\d_i^\dagger
  \d_i}{\Psi_0}$. We can write $\rho_0$ in terms of
$\{\d_i,\d_i^\dagger\}$. Then according to Wick's theorem, and the fact
$\ev{\c_i \c_j} = \ev{\d_n \d_m} = 0$, only terms of $\hat{O}_i \in \{\d_n
\d_n^\dagger,\d_n^\dagger \d_n \}$ survive. Then it is not hard to see
that $\rho_s$ can be represented as a product:

\beq
\rho_s = \prod_{i=1}^L \left(\meanvalue{\Psi_0}{\d_i
  \d_i^\dagger}{\Psi_0} \d_i \d_i^\dagger +
\meanvalue{\Psi_0}{\d_i^\dagger \d_i}{\Psi_0} \d_i^\dagger \d_i\right)
\label{eqn.rdm}
\eeq

%%%%%%%%%%%%%%%%%%%%%%%%%%%%%%%%

\subsubsection{Closed Form for the  Entanglement Entropy}
According to equation \ref{eqn.rdm}, we see

\beq
\S(\rho_s) = - \sum_{i}
  \left( \meanvalue{\Psi_0}{\d_i
  \d_i^\dagger}{\Psi_0} \ln(\meanvalue{\Psi_0}{\d_i
  \d_i^\dagger}{\Psi_0}) + \meanvalue{\Psi_0}{\d_i^\dagger \d_i}{\Psi_0}
  \ln (\meanvalue{\Psi_0}{\d_i^\dagger \d_i}{\Psi_0})\right)
\eeq

Follow the work of Jin and Korepin\cite{Jin2004}, we introduce
Majorana operators of $\c_i$s' and $\d_i$s'

\beq
\label{eqn:majorana}
\begin{split}
 c_{2l-1} &= \c_l+\c_l^\dagger =
  \frac{1}{\sqrt{2\pi}}\int_{-\pi}^{\pi} dk e^{-ikl}(\tilde{c}_k
  + \tilde{c}_k^\dagger) \\
 c_{2l} &= -i(\c_l - \c_l^\dagger) = \frac{1}{\sqrt{2
  \pi}}\int_{-\pi}^{\pi} dk e^{-ikn}i(\tilde{c}_k^\dagger - \tilde{c}_k)\\
 d_{2l-1} &= \d_l+\d_l^\dagger \text{  and  } d_{2l} = -i(\d_l -
\d_l^\dagger).\\
\end{split}
\eeq

They satisfy:

\beq
c_l = c_l^\dagger,\ \{c_l, c_m \} = 2 \delta_{lm},\ \ev{c_l} = 0.
\eeq

The two-point correlation functions now are:

\beq
\label{eqn.mjr.crlt}
\begin{split}
\ev{c_{2l-1} c_{2m-1}} & = \frac{1}{2\pi}\int_{-\pi}^{\pi} dk_1 dk_2
e^{-ik_1l + ik_2m}\ev{(\tilde{c}_{k_1}+\tilde{c}_{k_1}^\dagger)(\tilde{c}_{k_2}+\tilde{c}_{k_2}^\dagger)}\\
&= \frac{1}{2\pi}\int_{-\pi}^{\pi} dk e^{-i(l-m)k} = \delta_{lm}
= \ev{c_{2l} c_{2m}}\\
\ev{c_{2l-1} c_{2m}} & = \frac{1}{2\pi}\int_{-\pi}^{\pi} dk_1 dk_2
e^{-ik_1l + ik_2m}\ev{(\tilde{c}_{k_1}+\tilde{c}_{k_1}^\dagger) (-i)
  (\tilde{c}_{k_2}-\tilde{c}_{k_2}^\dagger)} \\
& = \frac{-i}{2 \pi} \int_{\pi}^{\pi} dk e^{-ik(l-m)}g(k) =
-\ev{c_{2l} c_{2m-1}}\\
g(k) &= \begin{cases} 1& \text{if $k > k_F$ or $k < -k_F$},\\ -1 &
  \text{if $k_F > k > -k_F$} \\ \end{cases}.
\end{split}
\eeq

So we have:

\beq
\label{eqn.mjr.mtx}
\ev{c_l c_m} = \delta_{lm} + i (\mathbf{B}_L)_{lm},
\eeq

here $\mathbf{B}_L = \mathbf{G}_L \otimes \begin{pmatrix} 0 & 1 \\ -1
  & 0\\ \end{pmatrix}$, with
$\mathbf{G}_L = \begin{pmatrix}g_0 &  g_{-1} & \dots  & g_{1-L} \\
g_1 & g_0 & \  & \vdots  \\
\vdots & \  & \ddots & \vdots \\
g_{L-1} & \dots & \dots & g_0 \\
\end{pmatrix}$

where $g_l$ is defined as:

\beq
g_l = \frac{1}{2\pi} \int_{-\pi}^{\pi}dk e^{-ilk}g(k), \
g(k) = \begin{cases} 1 & \text{if $k > k_F$ or $k < -k_F$},\\ -1 &
  \text{if $k_F > k > -k_F$} \\ \end{cases}
\eeq

As we found, once we diagonalize the correlation-function matrix, the
entanglement entropy is then given by $- \sum_i v_i \ln v_i $, where
$v_i$ is the i\textit{th} eigenvalue of the matrix. However, due to the
convention we adopted for the Majorana operators, the transformation
is not unitary, but with an extra factor of 2. Thus, in order to ensure
the normalization condition of the density matrix, we must divide this
factor of 2. Then the closed form for the entanglement entropy is
readily given as:

\beq
\label{eqn.entropy}
S(\rho_s) = \sum_{i=1}^L e(1,\nu_i)
\eeq

with

\beq
e(x,\nu) = -\frac{x+\nu}{2} \ln(\frac{x+\nu}{2}) - \frac{x-\nu}{2}
\ln(\frac{x-\nu}{2}), \text{  $\nu$' s are eigenvalues of $\mathbf{G}_L$}.
\eeq

However, to obtain all eigenvalues of $\nu_m$ directly from the matrix
$\mathbf{G}_L$ is nontrivial task. Let us introduce

\beq
\label{eqn.det}
D_L(\lambda) = \det(\tilde{\mathbf{G}}_L(\lambda)) \equiv \det(\lambda I_L-\mathbf{G}_L)
\eeq

$I_L$ is a identity matrix of dimension L. It is known that
  $\mathbf{G}_L$ is a Toeplitz matrix (see \cite{toeplitz_book}),
  i.e. its matrix elements depend solely on the difference between the
  two indices. Obviously we also have

\beq
D_L(\lambda) = \prod_{m=1}^L(\lambda - \nu_m).
\eeq

From the Cauchy residue theorem and the analytical property of
$e(x,\nu)$, $S(\rho_s)$ can be rewritten as

\beq
S(\rho_s) = \lim_{\epsilon \rightarrow 0^+} \lim_{\delta \rightarrow
  0^+} \frac{1}{2\pi i} \oint_{c(\epsilon,\delta)} e(1+\epsilon) d \ln D_L(\lambda).
\eeq

\begin{figure}
  \centering
  \includegraphics[height=15cm,angle=-90]{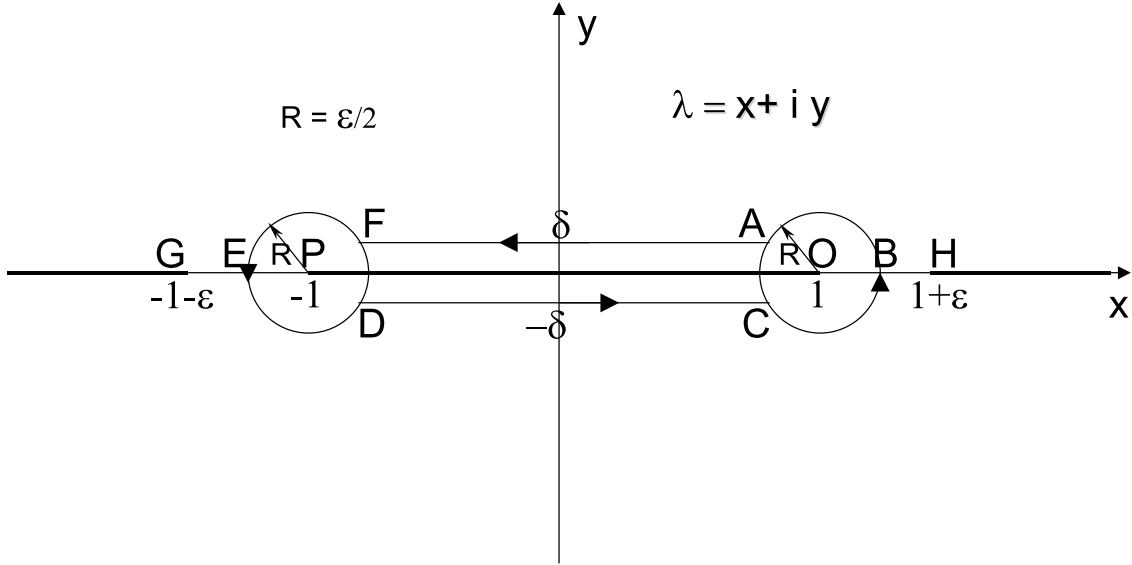}
  \caption{The contour $c(\epsilon,\delta)$. Bold lines
  ($-\infty,-1-\epsilon$) and ($1+\epsilon,\infty$) are the cuts of
  integrand $e(1+\epsilon,\lambda)$. Zeros of $D_L(\lambda)$
  [\ref{eqn.det}] are located on bold line (-1,1) and this line
  becomes the cut of $\text{d} \log D_L(\lambda)$ for $L \rightarrow
  \infty$. The arrow is the direction of the route of the integral we
  take and \textbf{R} is the radius of the circle.}
\end{figure}

Here the contour $c(\epsilon,\delta)$ in Fig.1 encircles all zeros of
$D_L(\lambda)$, but the function $e(1+\epsilon,\lambda)$ is analytic
within the contour. The Toeplitz matrix $\tilde{G}_L(\lambda)$ is
generated by the function $\tilde{g}(k)$ defined by

\beq
\tilde{g}(k) = \begin{cases} \lambda - 1 & -k_F < k < k_F \\
\lambda + 1 & k_F < k < (2\pi - k_F) \\\end{cases}
\eeq

Once the determinant of the Toeplitz matrix $\mathbf{\tilde{G}}_L$ is
obtained analytically, one will be able to get a closed analytic
result for $S(\rho_s)$.

%%%%%%%%%%%%%%%%%%%%%%%%%%%%%%%%%%%%%%%%%%%%%%%
\subsubsection{The Toeplitz Matrix and the Fisher-Hartwig Conjecture}

Toeplitz matrix \mbox{$T_{\mathrm{L}}[\phi]$} is said to be
generated by function \mbox{$\phi(\theta)$} if
\begin{eqnarray}
T_{\mathrm{L}}[\phi]= (\phi_{i-j}),~~~i,j=1,\cdots,\mathrm{L}-1
\end{eqnarray}
where
\begin{eqnarray}
\phi_l=\frac{1}{2\pi} \int_0^{2\pi} \phi (\theta) e^{-\mathrm{i} l \theta} \mathrm{d} \theta
\end{eqnarray}
is the $l$-th Fourier coefficient of generating function
\mbox{$\phi(\theta)$}. The determinant of
\mbox{$T_{\mathrm{L}}[\phi]$} is denoted by $D_{\mathrm{L}}$.

\noindent {\bf Fisher-Hartwig Conjecture:} Suppose the generating
function of Toeplitz matrix $\phi (\theta)$ is singular in the
following form
\begin{equation}
\phi(\theta)=\psi(\theta) \prod_{r=1}^R t_{\beta_r,\,
\theta_r}(\theta) u_{\alpha_r,\,\theta_r}(\theta)
\end{equation}
where
\begin{eqnarray}
t_{\beta_r,\,\theta_r}(\theta)&=&\exp [-i\beta_r (\pi-\theta+\theta_r)], \qquad \theta_r<\theta <2\pi+\theta_r\label{tdf}\\
u_{\alpha_r,\,\theta_r}(\theta)&=&\Bigl(2-2\cos
(\theta-\theta_r)\Bigr)^{\alpha_r},\quad\quad \Re
\alpha_r>-\frac{1}{2}\label{udf}
\end{eqnarray}
and $\psi$: $\mathbf{T}\to \mathbf{C}$ is a smooth non-vanishing
function with zero winding number. Then as $n\to \infty$, the
determinant of $T_{\mathrm{L}}[\phi]$
\begin{eqnarray}
D_{\mathrm{L}}= \left({\cal F}[\psi]\right)^{\mathrm{L}}
\left(\prod_{i=1}^R {\mathrm{L}}^{\alpha_i^2-\beta_i^2}\right)
{\cal E}[\psi, \{\alpha_i\}, \{\beta_i\},\{\theta_i\}],
~~\mathrm{L}\to \infty.\label{fh}
\end{eqnarray}
Here ${\cal F}[\psi]=\exp \left(\frac{1}{2\pi} \int_0^{2\pi}\ln
\psi(\theta) \mathrm{d} \theta\right)$.
 Further assuming that there exists Weiner-Hopf factorization
\begin{equation}
\psi(\theta)= {\cal F}[\psi]\,
\psi_+\Bigl(\exp(\mathrm{i}\theta)\Bigr)\,
\psi_-\Bigl(\exp(-\mathrm{i}\theta)\Bigr),
\end{equation}
then constant \mbox{${\cal E}[\psi, \{\alpha_i\},
\{\beta_i\},\{\theta_i\}]$} in Eq.~\ref{fh} can be written as
\begin{eqnarray}
{\cal E}[\psi, \{\alpha_i\},\{\beta_i\},\{\theta_i\}]&=&{\cal E}[\psi]
\prod_{i=1}^R G(1+\alpha_i+\beta_i) G(1+\alpha_i-\beta_i)/G(1+2\alpha_i) \nonumber\\
&\times &\prod_{i=1}^R \biggl(\psi_-\Bigl(\exp(\mathrm{i}
\theta_i)\Bigr)\biggr)^{-\alpha_i-\beta_i} \biggl(\psi_+
\Bigl(\exp(- \mathrm{i} \theta_i)\Bigr)\biggr)^{-\alpha_i+\beta_i}\nonumber\\
&\times& \prod_{1\leq i \neq j \leq R}
\biggl(1-\exp\Bigl(\mathrm{i}
(\theta_i-\theta_j)\Bigr)\biggr)^{-(\alpha_i+\beta_i)(\alpha_j-
\beta_j)},
\end{eqnarray}
$G$ is the Barnes $G$-function, \mbox{${\cal
E}[\psi]=\exp(\sum_{k=1}^{\infty} k s_k s_{-k})$}, and $s_k$ is
the $k$-th Fourier coefficient of \mbox{$\ln \psi(\theta) $}. The
Barnes $G$-function is defined as
\begin{eqnarray}
G(1+z)=(2\pi)^{z/2} e^{-(z+1)z/2-\gamma_E z^2/2}
\prod_{n=1}^{\infty} \{ (1+z/n)^n e^{-z+z^2/(2n)}\},
\end{eqnarray}
where $\gamma_E$ is Euler constant and its numerical value is
\mbox{$0.5772156649\cdots$}. This conjecture has not been proven
for general case. However, there are various special cases for
which the conjecture was proven.

%%%%%%%%%%%%%%%

For our case, the generating function $\tilde{g}(\theta)$ has two
jumps at $\theta=\pm k_F$ and it has the following canonical
factorization
\begin{eqnarray}
\tilde{g}(\theta)= \psi(\theta)
t_{\beta_1(\lambda),\,k_F}(\theta)t_{\beta_2(\lambda),\,
-k_F}(\theta) \label{bbta1}
\end{eqnarray}
 with
\begin{eqnarray}
\psi(\theta)=(\lambda+1)\left(\frac{\lambda+1}{\lambda-1}\right)^{-k_F/\pi},
~~~\beta(\lambda)=
-\beta_1(\lambda)=\beta_2(\lambda)=\frac{1}{2\pi \mathrm{i}} \ln
\frac{\lambda+1}{\lambda-1}. \label{bbta2}
\end{eqnarray}
The function $t$ was defined in Eq.~$\ref{tdf}$. We fix the branch
of the logarithm in the following way
\begin{eqnarray}
 -\pi \leq \arg \left(\frac{\lambda+1}{\lambda-1}\right) < \pi. \label{bbta3}
\end{eqnarray}

Here we verify the the above factorization expression
explicitly. We have two jumps at $\theta = \pm k_F$.  According to Eq.33,
we have two t function:

\beq
\begin{split}
t_{\beta_1,k_F} &= \exp[-i\beta_1(\pi - \theta + k_F)],\ k_F < \theta
<2\pi + k_F\\
t_{\beta_2,-k_F} &= \exp[-i\beta_2(\pi - \theta - k_F)], \ -k_F <\theta <
2\pi - k_F\\
\end{split}
\eeq

Then $\tilde{g}(\theta)$ is given as:

\beq
\begin{split}
\tilde{g}(\theta) &= (\lambda + 1)\left(\frac{\lambda + 1}{\lambda -
  1}\right)^{-k_F/\pi} \exp[-i(\frac{1}{2\pi i})\ln (\frac{\lambda +
  1}{\lambda - 1}) (\pi - \theta - k_F)] \\ & \times \exp[-i
  (\frac{-1}{2\pi i}) \ln (\frac{\lambda + 1}{\lambda - 1})(\pi -
  \theta + k_F)]\\
& = (\lambda + 1)\left(\frac{\lambda + 1}{\lambda - 1}\right)^{ -k_F /
  \pi} \exp[-\frac{\pi - \theta -k_F}{2 \pi}\ln (\frac{\lambda +
  1}{\lambda - 1})] \exp[\frac{\pi - \theta + k_F}{2\pi} \ln
  (\frac{\lambda + 1}{\lambda - 1})]
\end{split}
\eeq

When $\theta \in [-k_F,k_F]$, $$t_{\beta_2,-k_F} = (\frac{\lambda +
  1}{\lambda - 1})^{-\frac{\pi - \theta - k_F}{2\pi}}$$. However, for
  $t_{\beta_,k_F}$, $\theta$ is not defined in this region, but
  considering periodicity, $\theta \in [2\pi-k_F,2\pi+k_F]$, and to
  move it into the same region as of $t_{\beta_2,-k_F}$, we must write
  $t_{\beta_1,k_F}$ as $$t_{\beta_1,k_F} = (\frac{\lambda + 1}{\lambda
  - 1})^{\frac{\pi - (\theta + 2\pi) + k_F}{2\pi}} = (\frac{\lambda + 1}{\lambda
  - 1})^{\frac{k_F - \theta - \pi}{2\pi}}$$

Then we have

\beq
\tilde{g}(\theta) = (\lambda + 1)(\frac{\lambda + 1}{\lambda -
  1})^{-1} = \lambda - 1,\ \text{when $\theta \in [-k_F,k_F]$}
\eeq

Similar argument also applies to $\theta \in [k_F,2\pi-k_F]$. And we
have verified that the factorization works.

For $\lambda \notin [-1,1]$, we know that $|\Re (\beta_1(\lambda))
|<\frac{1}{2}$ and $| \Re (\beta_2(\lambda)) | < \frac{1}{2}$ and
Fisher-Hartwig conjecture was {\bf PROVEN} by E.L.\ Basor for this
case \cite{basor}. Therefore, we will call it the theorem instead
of conjecture for our application.  Hence following the theorem in
Eq.~$\ref{fh}$, the determinant $D_{\mathrm{L}}(\lambda)$ of
$\lambda I_{\mathrm{L}} -\mathbf{G}_{\mathrm{L}}$ can be
asymptotically represented as
\begin{eqnarray}
D_{\mathrm{L}}(\lambda)&=&\Bigl(2-2\cos(2
k_F)\Bigr)^{-\beta^2(\lambda)}
\left\{G\Bigl(1+\beta(\lambda)\Bigr) G\Bigl(1-\beta(\lambda)\Bigr)\right\}^2\nonumber \\
&&\left\{(\lambda+1)\Bigl((\lambda+1)/(\lambda-1)\Bigr)^{-k_F/\pi}\right\}^{\mathrm{L}}
\mathrm{L}^{-2 \beta^2(\lambda)}.\label{apd}
\end{eqnarray}
Here $\mathrm{L}$ is the length of sub-system A and $G$ is the
Barnes $G$-function and
\begin{eqnarray}
G(1+\beta(\lambda))
G(1-\beta(\lambda))=e^{-(1+\gamma_E)\beta^2(\lambda)}\prod_{n=1}^{\infty}
 \left\{\left(1-\frac{\beta^2(\lambda)}{n^2}\right)^n e^{\beta^2(\lambda)/n^2} \right\}.
\end{eqnarray}

Therefore,

\beq
\begin{split}
\ln D_L(\lambda) &= L(\ln(\lambda + 1) -\frac{k_F}{\pi}\ln(\lambda + 1)
+\frac{k_F}{\pi}\ln(\lambda - 1)) - 2 \beta^2(\lambda) \ln L \\
& + 2(\ln (G(1 + \beta)) + \ln(G(1 - \beta))) -\beta^2(\lambda)\ln(2 - 2
\cos 2k_F) \\
\end{split}
\eeq

\beq
\begin{split}
\frac{\partial{\ln D_L}}{\partial{\lambda}} &=
L(\frac{1-k_F/\pi}{\lambda + 1} + \frac{k_F/\pi}{\lambda - 1}) -
\frac{\partial{\beta(\lambda)}}{\partial{\lambda}}(4 \beta(\lambda)
\ln L \\ &- 2 (\frac{G'(1+\beta)}{G(1+\beta)} + \frac{G'(1 - \beta)}{G(1 -
  \beta)}) + 2 \beta(\lambda) \ln(2 - 2 \cos k_F)) \\
\end{split}
\eeq

Here our only concern is the first term and the second term which
diverge linearly or logarithmically as the size the subsystem.

%%%%%%%%%%%%%%%%%%%%%%%%%%%%%%%%%%

\subsubsection{Asymptotic Behavior of the Entanglement Entropy}
Now let us proceed to calculate the leading order of the entanglement
entropy according our results.

First, we consider the term grows as subsystem size L. It is not
difficult to see that the contribution from this term is actually
zero, since the only residues arising from poles at $\lambda = \pm 1$
are just zero.

\beq
\begin{split}
S_1(\rho_s) &= \lim_{\epsilon \rightarrow 0^+} \lim_{\delta \rightarrow 0^+}
\frac{1}{2\pi i} \oint_c \text{d} \lambda \left(-\frac{1+\epsilon
  +\lambda}{2} \ln (\frac{1+\epsilon + \lambda}{2}) -\frac{1+\epsilon
  - \lambda}{2} \ln (\frac{1+\epsilon - \lambda}{2}) \right) \\
& \times (\frac{1 - k_F/\pi}{1 + \lambda} - \frac{k_F/\pi}{1 -
  \lambda}) L\\
& = 0 \\
\end{split}
\eeq

Now let us turn to the second leading term of order $\ln L$

\beq
\begin{split}
S_2 &= \frac{2}{\pi^2} \oint \text{d} \lambda \frac{e(1+\epsilon,
  \lambda) \beta(\lambda)}{(1 + \lambda)(1 - \lambda)} \ln L\\
& = \frac{2}{\pi^2} \oint \text{d} \lambda \left(-\frac{1+\epsilon
  +\lambda}{2} \ln (\frac{1+\epsilon + \lambda}{2}) -\frac{1+\epsilon
  - \lambda}{2} \ln (\frac{1+\epsilon - \lambda}{2}) \right) \\
& \times  \frac{1}{2 \pi i} \ln \left(\frac{\lambda + 1}{\lambda -
  1}\right) \frac{\ln L}{(1 + \lambda)(1 - \lambda)} \\
\end{split}
\eeq

This contour integral can be calculated as follows. First, let us look
at Fig.1, noting that

\beq
\oint_{c(\epsilon,\delta)} \text{d}\lambda(\dots) =
\left(\int_{\overrightarrow{AF}} + \int_{\overrightarrow{FED}} +
\int_{\overrightarrow{DC}} + \int_{\overrightarrow{CBA}} \right) \text{d} \lambda(\dots),
\eeq

according to contour integral theory, this contour yields the same
result as

\beq
\oint_{c(\epsilon,\delta)} \text{d}\lambda(\dots) =
\left((\int_{\overrightarrow{AF}} + \int_{\overrightarrow{FD}} +
\int_{\overrightarrow{DC}} + \int_{\overrightarrow{CA}}) +
(\int_{\overrightarrow{FEDF}} + \int_{\overrightarrow{ACBA}} ) \right)
\text{d} \lambda(\dots) .
\eeq

The contour $\int_{\overrightarrow{FEDF}}$ and
$\int_{\overrightarrow{ACBA}}$ are merely closed circles around the
points $\pm 1$. As we take the limit $\epsilon \rightarrow 0^+$, they
shrink and contain only this two points. Using Cauchy's residue
theorem, it is easy to see that the contribution from these two
contours is zero.

So the contour integral is simplified to

\beq
S_2 = \frac{2}{\pi^2}\left(\int_{-1 + i0^+}^{1 + i0^+} - \int_{-1 +
  i0^-}^{1 + i0^-}\right) \text{d} \lambda \frac{e(1+\epsilon,
  \lambda) \beta(\lambda)}{(1 + \lambda)(1 - \lambda)} \ln L.
\eeq

The rest part of the integral function is analytic within the contour,
however, $\beta (\lambda)$ could have jumps in its angular part since
we fix the branch by requiring $-\pi \leq arg\left( \ln \frac{\lambda +
  1}{\lambda - 1}\right)$.

\beq
\begin{split}
\beta(x + i0^\pm) &= \frac{1}{2 i \pi} \ln \left(\frac{x + i0^\pm +
  1}{x + i0^\pm - 1}\right) \\
& = \frac{1}{2i\pi}\left(\ln\frac{1 + x}{1 - x} + \ln \frac{1 +
  \frac{i0^\pm}{1 + x}}{-1 + \frac{i0^\pm}{1 - x}}\right) \\
& \simeq \frac{1}{2i\pi}\left(\ln\frac{1 + x}{1 - x} + \ln(-(1 + i0^\pm))
  \right) \\
& \simeq \frac{1}{2i\pi}\left(\ln\frac{1 + x}{1 - x} + \ln(e^{i(\pi + 0^\pm)})
  \right).
\end{split}
\eeq

Taking into account the branch-cut condition, one immediately see that

\beq
\begin{split}
& \ln e^{i(\pi + 0^+)} = -i(\pi - 0^+) \\
& \ln e^{i(\pi + 0^-)} = i(\pi -0^-) \\
\end{split}
\eeq

Therefore, our contour integral can now be written as

\beq
S_2 = \frac{2}{\pi^2}\int_{-1}^1 \text{d} \lambda
\frac{e(1,\lambda)}{(1+\lambda)(1-\lambda)} \ln L = \frac{1}{3} \ln L
\eeq

Thus we have obtained the leading order behavior of the entanglement
entropy of a segment of length $L$ embedded in an infinite spin-less
free fermion lattice. Here we did not calculate the sub-leading terms,
however, when $k_F$, the Fermi momentum or filling factor, becomes
extremely small or very close to 1, they will become more and more
important and eventually kill the entanglement's logarithm dependence
on L. The criterion is given by $\mathcal{L} = 2L\sin k_F \gg 1$.

\subsection{Bosonic Approach}                              %
%%%%%%%%%%%%%%%%%%%%%%%%%%%%%%%%%%%%%%%%%%%%%%%%%%%%%%%%%%%%%%%%%%%%%%%%%

%%%%%%%%%%%%%%%%
\subsubsection{Brief Introduction to Bosonization of 1D Spinless Fermion}
In this section, we shall briefly introduce the bosonization of
fermionic systems in one dimension. We will generally follow
\cite{delft1998} which follows Haldane's constructive approach.

For simplicity we will only consider the bosonization of a theory
involving only one species of fermions. And bosonization of a theory
is possible whenever the following prerequisites are met:
\begin{enumerate}
\item{The theory can be formulated in terms of a set of fermion
  creation and annihilation operators with canonical anti-commutation
  relations: $\{c_k,c_{k'}^\dagger\} = \delta_{kk'},\ k \in [-\infty,
  \infty]$;}
\item{The label $k$ above is a discrete, unbounded momentum index of
  the form $k = \frac{2\pi}{L}(n_k - \frac{1}{2}\delta_b)$ with $n_k
  \in \mathbb{Z}$ and $\delta_b \in [0,2)$. Here $n_k$ are integers, $L$
  is the length of the system size, and $\delta_b$ is a parameter
  that will determine the boundary condition for the fermion fields
  defined below.}
\end{enumerate}
Obviously the prerequisites here are not directly satisfied by the
lattice free fermion model we considered in previous
sections. However, they can be satisfied by doing the following
procedures: i)performing a particle-hole transformation $c^\dagger_k
\rightarrow \tilde{c}_k$ for $k < -k_F$, ii) then shifting the Fermi
point to $k_F = 0$, iii) letting the lattice spacing $a \rightarrow
0$, iv) extending definition of $k$ to $(-\infty, \infty)$.

{\it The fermion fields}:

\beqarr
\psi (x) &\equiv& (\frac{2\pi}{L})^{1/2} \sum_{k = - \infty}^{\infty}
e^{-ikx} c_k,\\
\text{with inverse } c_k &=& (2\pi L)^{-1/2}\int_{-L/2}^{L/2} dx
e^{ikx} \psi (x).
\eeqarr

And given a set of discrete $k$'s of the form above, the fields
$\psi(x)$ obey the following periodicity condition:

\beq
\psi (x+L/2) = e^{i\pi \delta_b} \psi (x-L/2).
\eeq

Using the following identity\cite{shilov64}

\beq
\sum_{n \in \mathbb{Z}} e^{iny} = 2\pi \sum_{\bar{n} \in
  \mathbb{Z}}\delta(y-2\pi \bar{n})
\eeq
we can immediately get the anti-commutation relations:

\beqarr
\{\psi(x),\psi^\dagger (x')\} &=& \frac{2\pi}{L} \sum_{n \in \mathbb{Z}}
e^{-i(x-x')(n-\delta_b/2) 2 \pi / L} = 2\pi \sum_{\bar{n} \in
  \mathbb{Z}}\delta(x-x' - \bar{n} L)e^{i\pi \bar{n} \delta_b};\\
\{\psi(x),\psi (x')\} &=& 0.
\eeqarr

{\it Vacuum State} $\ket{0}_0$:

Let $\ket{0}_0$ be the state defined by the properties

\beqarr
c_k \ket{0}_0 \equiv 0 & \text{for } & k>0, \label{vacuum1}\\
c^\dagger_k \ket{0}_0 \equiv 0 & \text{for } & k \leq 0. \label{vacuum2}
\eeqarr
 We shall call $\ket{0}_0$ the {\it vacuum state} and use it as
 reference state relative to which the occupations of all other states
 in Fock space are specified. With this definition we can define the
 operation of {\it fermion-normal-ordering}, to be denoted by $\nmod{\
 }$, with respect to this vacuum state: to fermion-normal-order a
 function of $c$ and $c^\dagger$'s, all $c_k$ with $k > 0$ and all
 $c_k^\dagger$ with $k \leq 0$ are to be moved to the right of all
 other operators(i.e. all all $c_k^\dagger$ with $k > 0$ and all
 $c_k$ with $k \leq 0$), so that:

\beq
\nmod{ABC\dots} = ABC\dots -
_0\meanvalue{0}{ABC\dots}{0}_0 \quad \text{for } A,B,C,\dots
\in \{c_k;c_k^\dagger\}
\eeq

$\vec{N}${\it -particle ground state} $\ket{\vec{N}}_0$:
Let $\hat{N}$ be the operator that counts the number of electrons
relative to $\ket{0}_0$:
\beq
\hat{N} \equiv \sum_{k = - \infty}^\infty \nmod{ c^\dagger_k c_k} =
\sum_{k = - \infty}^{\infty} [c^\dagger_k c_k - {_0}
  \meanvalue{0}{c^\dagger_k c_k}{0}_0 ]
\eeq

The set of all states with the same $\hat{N}$-eigenvalues $N$ will be
called the $N$-particle Hilbert space $H_N$. It contains infinite
number of states, corresponding to different particle-hole
excitations. Let us denote all of them by $\ket{N}$. For a given $N$,
there is a state which contains no particle-hole excitations. We will
denote it as $\ket{N}_0$. To prevent possible ambiguities in its
phase, we define it by the following order:

\beq
\ket{N}_0 \equiv
\begin{cases}
c^\dagger_N c^\dagger_{(N-1)}\dots c^\dagger_1 \ket{0}_0  &\text{for } N>0,\\
 \ket{0}_0  & \text{for } N = 0\\
c_{N+1} c_{(N+2)}\dots c_0 \ket{0}_0 &\text{for } N < 0.\\
\end{cases}
\eeq

{\it Bosonic operators} $b^\dagger_q$ {\it and} $b_q$:
\beq\label{N0}
b^\dagger_q \equiv \frac{i}{\sqrt{n_q}} \sum_{k=-\infty}^{\infty}
c^\dagger_{k+q} c_k, \quad b_q \equiv \frac{-i}{\sqrt{n_q}}
\sum_{k=-\infty}^{\infty} c^\dagger_{k-q} c_k,
\eeq
with $q \equiv \frac{2\pi}{L} n_q > 0$ where $n_q \in \mathbb{Z}^+$ is
a positive integer. Thus the bosonic creation and annihilation
operators are defined for $q > 0$ only.

And it is not hard to prove the following bosonic commutation
relations:

\beqarr
[b_q, b_{q'}] &=& [b^\dagger_q, b^\dagger_{q'}] = 0,  [N_q, b_{q'}]
= [N_q, b^\dagger_{k'} = 0],\quad  \text{for all } q,\ q' ;\\
\ [b_q, b_{q'} ^\dagger] &=& \sum_{k=-\infty}^\infty \frac{1}{n_q}
(c^\dagger_{k+q-q'} c_k - c^\dagger_{k+q} c_{k+q'}) = \delta_{qq'}.
\eeqarr

Making a connection with Eq.(\ref{N0}), we could see that in each
$N$-particle Hilbert space $H_N$, $\ket{N}_0$ functions as vacuum
state for bosonic operators defined above:

\beq
b_q \ket{N}_0 = 0, \quad \text{for all} \quad k.
\eeq

This is because $\ket{N}_0$ is the $N$-particle ground state and does
not contain any particle-hole excitations(bosonic excitations).

With a proper construction of a set of bosonic operators in $k$ space,
the construction of boson fields is just straightforward:
\beq
\varphi (x) \equiv -\sum_{q>0} \frac{1}{\sqrt{n_q}}e^{-iqx} b_q
e^{-aq/2}, \quad \varphi ^\dagger (x) \equiv -\sum_{q>0}
\frac{1}{\sqrt{n_q}} e^{iqx} b^\dagger_q e^{-aq/2}.
\eeq

Here $a>0$ is an infinitesimal regularization parameter which is used
to regularize $q \rightarrow \infty$ divergent momentum sums that
arise in certain non-normal-ordered expressions and commutators. The
following commutation relations can be verified for the boson fields
defined above:

\beqarr
[\varphi(x), \varphi(x')] &=& [\varphi^\dagger (x), \varphi^\dagger
  (x')] = 0,\\
\ [\varphi(x), \varphi^\dagger (x')] &=& \sum_{q > 0}
\frac{1}{n_q}e^{-q[i(x-x')+a]} = - \ln[1-e^{-i\frac{2 \pi}{L}(x - x' -
    ia)}]
\eeqarr

Then consider their Hermitian combination:

\beq
\phi(x) \equiv \varphi(x) + \varphi^\dagger (x).
\eeq

Check the canonical commutation relation:

\begin{equation}
\begin{split}
& [\phi(x), \partial_{x'}\phi(x')] =
i\frac{2\pi}{L}\left(\frac{1}{e^{i\frac{2\pi}{L}(x-x'-ia)} -1} +
\frac{1}{e^{-i\frac{2\pi}{L}(x-x'+ ia)} -1}\right)\\
& \xrightarrow{L \rightarrow \infty}  2\pi i[\frac{a/\pi}{(x-x')^2 +
    a^2}-\frac{1}{L}] \xrightarrow{a \rightarrow 0} 2\pi i[\delta(x-x')-\frac{1}{L}],\\
\end{split}
\end{equation}

which is exactly the canonical commutation relation for boson fields
when $L \rightarrow \infty$.

{\it Bosonization and entanglement entropy}

Let us look at the corresponding mapping in real space since the
partition of the system is usually carried out in real
space. Therefore, to legitimate our use of bosonization to study
entanglement entropy, we have to establish the relation between the
fermion fields and the boson fields in real space and show that the
mapping, at least approximately, preserve the partition of the system.

First, let us look at the fields:

\beq
\begin{split}
 \varphi(x) &= \sum_{q > 0} \frac{1}{\sqrt{n_q}}e^{-iqx} e^{-aq/2}
\frac{-i}{\sqrt{n_q}} \sum_{k=-\infty}^{\infty} c^\dagger_{k-q} c_k\\
& = \sum_{q>0}\frac{-i}{n_q} e^{-aq/2} e^{-iqx} \sum_k \frac{1}{2\pi
L} \int^{L/2}_{-L/2} dx_1 e^{-i(k-q)x_1}\psi^\dagger(x)
\int^{L/2}_{-L/2}
 e^{ikx_2}\psi(x_2)\\
 &= \int dx_1 dx_2 \sum_{q>0} \frac{-i}{2 \pi L n_q} \psi^\dagger(x_1)
\psi(x_2) \sum_{k} e^{i(-aq/2 -qx + qx_1 - kx_1 + kx_2)}\\
& = \int dx_1 \sum_{n_q > 0} \frac{-i}{L n_q}\psi^\dagger(x_1)
\psi(x_1)
e^{i\frac{2\pi n_q}{L}(x_1-x -a/2)}\\
& = \int dx_1 \frac{i}{L} \psi^\dagger(x_1)\psi(x_1) \ln(1 -
e^{i\frac{2\pi}{L}(x_1 - x - a/2)}).
\end{split}
\eeq

It seems that the mapping of fields does not fulfill our
requirement. However, if we look at the fermion density operator which is
the object one will directly work with when calculating entanglement
entropy, we shall see that our requirement is indeed satisfied.

\beq
\begin{split}
\rho_f(x) &\equiv \nmod{\psi^\dagger(x) \psi(x)} = \frac{2\pi}{L}
\sum_{q} e^{-iqx}\sum_{x} \nmod{c^\dagger_{k-q} c_k}\\
& =  \frac{2\pi}{L}\sum_{q>0}i \sqrt{n_q} (e^{-iqx} b_q - e^{iqx}
b_q) + \frac{2 \pi}{L} \sum_{k} \nmod{c_k^\dagger c_k} \\
& =  \partial_x \phi(x) + \frac{2\pi}{L} \hat{N}, \quad \text{for $a
  \rightarrow 0$}.
\end{split}
\eeq

For the whole system, the particle number is conserved, so $\hat{N}$
is just a number. Thus we have justified our utilization of bosonization to
study the entanglement entropy in many-fermion systems.

However, we still need to find out corresponding bosonic states of the
fermionic states we are interested in. At present we are only
interested in fermion ground state at zero temperature, i.e. the Fermi
sea. It is the $\ket{0}_0$ we defined in Eq.(\ref{vacuum1}) and
(\ref{vacuum2}) which corresponds to the vacuum state $\ket{N = 0}_0$ of
the boson modes.

\subsubsection{Entanglement Entropy of free Bosons}
In this part,we shall first introduce available analytic result in
lattice model. However, though this approach can give a nice analytic
expression for entanglement entropy, we can not get desired result
explicitly due to technique difficulty. Then we will give a short
introduction of the field theory approach, following Calabrese and
Cardy\cite{calabrese-2004-0406}.

%%%%%%%%%%%%%%%%%%%
{\it Lattice Model}

Let us consider a system of coupled harmonic oscillators in which the
Hamiltonian can generally be written in matrix form as:
\beq
H = \sum_{i,j}[T_{i,j}p_i p_j + V_{i,j}x_i x_j],
\eeq
where the operators $x_i$'s and $p_i$'s obey the canonical commutation
relation: $[x_i,p_j] = i \delta_{ij}$.
Consider its correlation functions of positions and momenta
\beq
\hat{X}_{n,m} = \ev{x_n x_m},\  \hat{P}_{n,m} = \ev{p_n p_m}.
\eeq

The ground state is a Gaussian state, and the multi-point correlation
functions observe Wick's theorem:
\beq
\ev{x_n x_m x_k x_l} = \ev{x_n x_m} \ev{x_k x_l} + \ev{x_n x_k}
\ev{x_m x_l} + \ev{x_n x_l} \ev{x_m x_k}.
\eeq

This also holds inside the subsystem when we do the
truncation. According to Wick's theorem, this indicates the density
matrix of the subsystem, i.e. the reduced density matrix is an
exponential of momenta and spatial coordinates. Therefore, in
principle we should be able to write the reduced density matrix as:
\beq
\rho_s = \mathcal{K}e^{-H'} = e^{-\sum_{i,j}[T_{i,j}'p_i p_j + V_{i,j}'x_i x_j]},
\eeq
here $\mathcal{K}$ is a normalization factor. It is in general not
easy to obtain an explicit analytic expression for $T'$ and
$V'$ due to two reasons: first, even in the simplest case of nearest
neighbor coupling, terms that do not conserve particle numbers would
arise; second, the transformation must be symplectic \footnote{
  "simplectic" here means preserving the commutator $[x_i,p_j] =
  \delta_{ij}$, i.e. a transformation that preserves the "symplectic"
  matrix $\begin{pmatrix}0 & 1 \\ -1 & 0 \end{pmatrix}$.}. However, for
such a Hamiltonian, it is always possible to find a symplectic
transformation $c_j = \frac{1}{\sqrt{2}}(x_j + i p_j), a_k = \sum_{j}
(A_k(j) c_j + B_k^\dagger(j) c^\dagger_j$ which can symplecticly
diagonalize the Hamiltonian to the following form:
\beq
H' = \sum_k \varepsilon(k) a_k^\dagger a_k.
\eeq

The eigenvalues $\varepsilon(k)$'s follow from the eigenvalues
$\nu_k^2$ of the $X'P'$\footnote{Here ' indicates that these are
  obtained by truncating the original matrices $X$ and $P$.} matrix via
\beq
\text{cth}(\varepsilon(k)/2) = \nu_k/2.
\eeq

And the entanglement entropy is given by

\beq
S = \sum_k ((\nu_k+1/2)\ln(\nu_k+1/2) - (\nu_k-1/2)\ln(\nu_k-1/2)).
\eeq
However, to obtain the full spectrum of the matrix $X'P'$ and calculate
the entanglement entropy is highly
non-trivial. No pure analytic approach has been developed, but with
the help of numerical methods, one still could reproduce desired
scaling law of the entanglement entropy. In the strong coupling limit
which corresponds to a massless free field, the scaling behavior has
been shown indeed to be $\frac{1}{3} \log L$.

%%%%%%%%%%%%%%%%%%%%%%%%%%%
{\it Field Theory Approach}\cite{calabrese-2004-0406}

Consider a quantum field theory in one dimension space and one time
dimension, described by the following action
\beq
S = \int \frac{1}{2}\left( - (\pi)^2 + m^2 \psi^2 \right)
d^2\tau = \int \frac{1}{2}\left( (\partial_\mu \psi)^2 + m^2 \psi^2
\right) d^2\tau,
\eeq
where $\pi(x) = i \partial_\mu \psi$ is the canonical momentum, and
satisfies the canonical commutation relation $[\psi(x),\pi(x')] = i \delta(x-x')$.
The density matrix $\rho$ in a thermal state at inverse temperature
$\beta$ is

\beq
\rho(\{\phi(x'')''\}|\{\phi(x')'\}) = Z(\beta)^{-1}
\meanvalue{\{\phi(x'')''\}}{e^{-H}}{\{\phi(x')'\}},
\eeq
where $Z(\beta) = \tr e^{-\beta H}$ is the partition function, and
$\{\phi(x)\}$ are the corresponding eigenstates of $\psi(x)$: $\psi(x)
\ket{\{\phi(x')\}} = \phi(x') \ket{\{\phi(x')\}} $.
This can be expressed as a (Euclidean) path integral:
\beq
\rho = Z^{-1} \int [d \phi(x,\tau)] \prod_x\delta(\phi(x,0)-\phi(x')')
\prod_x\delta(\phi(x,\beta) - \phi(x'')'') e^{-S_E},
\eeq
where $S_E = \int^\beta_0 L_E d\tau$, with $L_E$ the euclidean Lagrangian.
The normalization factor $Z$, i.e. the partition function is found by
setting $\{\psi(x)''\} = \{\psi(x)' \}$ and integrating over these
variables. This has the effect of sewing together the edges along
$\tau = 0$ and $\tau = \beta$ to form a cylinder of circumference
$\beta$.

The reduced density matrix of an interval $A = (x_i,x_f)$ can be
obtained by sewing together only those points which are not in the
interval $A$. This has the effect of leaving an open cut along the line
$\tau = 0$. For the calculation of entanglement entropy, we need to
perform a replica trick here. Instead of calculating $\tr \rho_A \log
\rho_A$, we compute $\rho_A^n$ first, for any positive integer $n$. To
do this, we make $n$ copies of above set-up labeled by an integer $k$
with $1 \le k \le n$, and sew them together cyclically along the open
cut so that $\psi(x)'_k = \psi(x)''_{k+1} (\text{and }\psi(x)'_n =
\psi(x)''_1)$ for all $x \in A$. Let us denote the path integral on
this $n$-sheeted structure by $Z_n(A)$. Then
\beq
\tr \rho_A^n = \frac{Z_n(A)}{Z^n},
\eeq
and
\beq
S_A = -\lim_{n \rightarrow 1} \frac{\partial}{\partial n}
\frac{Z_n(A)}{Z^n}.
\eeq

For a free theory on such n-sheeted geometry, it is easier to use the
identity\footnote{This only holds for non-interacting theories.}:

\beq
\frac{\partial}{\partial m^2} \log Z_n = -\frac{1}{2} \int G_n(\mathbf{r}
,\mathbf{r}) d^2r,
\eeq
where $G_n(\mathbf{r}, \mathbf{r'})$ is the Green's function in the
n-sheeted geometry. To obtain $\frac{Z_n(A)}{Z^n}$, we need $G_n - n
G_1$. The Green's function can be obtained by solving the Helmholtz
equation

\beq
(-\nabla^2_{\mathbf{r}} + m^2) G_n(\mathbf{r}, \mathbf{r'}) =
\delta^2(\mathbf{r-r'})
\eeq
with n-sheeted geometry. In polar coordinates(2D) this simply means extend
the domain of $\theta$ to $[0,2\pi n)$. However, an explicit formula
  is only available for infinite volume, which means we have to extend
  the subsystem to a semi-infinite one.

After solving for $G_n(\mathbf{r}, \mathbf{r'})$, let $\mathbf{r' =
  r}$ and perform the integral, we have

\beq
\frac{\partial}{\partial m^2} \log (\frac{Z_n}{Z^n}) = -\frac{1}{2}
\int (G_n(\mathbf{r,r}) - nG_1(\mathbf{r,r})) = \frac{1}{24
  m^2}(n-\frac{1}{n}).
\eeq

This will lead us to the final entanglement entropy

\beq
S = -\tr \rho \log \rho = -\frac{\partial}{\partial n} \tr
\rho^n|_{n=1} =- \frac{\partial}{\partial n}
(m^2a^2)^{\frac{1}{24}(n-1/n)}|_{n=1} = -\frac{1}{12} \log m^2 a^2.
\eeq
The power of $a$ is inserted to make the result dimensionless,
following Cardy's convention.
In the massless limit where $m \rightarrow 0$ which we are more
interested in, physically it is natural to replace $m$ with the
inverse subsystem size $L^{-1}$ since the coherent length diverges and
subsystem size is the only relevant characteristic size in this
problem. Also be careful that we actually dealt with a semi-infinite
subsystem which has only one boundary of partition in above
approach\footnote{This is verifiable in several cases.}. When we
consider a more common subsystem which has two boundaries of partition,
we want to double this entanglement entropy as it is a boundary
effect. Thus we recover the result $S = \frac{1}{3} \log L$ in
agreement with other approaches.

%\subsection{Conformal Field Theory Approach}

%%%%%%%%%%%%%%%%%%%%%%%%%%%%%%%%%%%%%%%%%%%%%%%%%%%%%%%%%%%%%%%%%
\section{Next Step: Generalization to Higher Dimensions}        %
%%%%%%%%%%%%%%%%%%%%%%%%%%%%%%%%%%%%%%%%%%%%%%%%%%%%%%%%%%%%%%%%%

%%%%%%%%%%%%%%%%%%%%
\subsection{Violation of Area Law in Fermionic Systems in Higher
  Dimensions: Known Results}
In Wolf's work\cite{wolf:010404}, he considers a general number
preserving quadratic Hamiltonian
\beq
H = \sum_{\alpha,\beta} T_{\alpha,\beta} c^\dagger_\alpha c_\beta, T = T^\dagger,
\eeq
describing Fermions on a d-dimensional cubic lattice, so that each
component of the vector $\alpha$, $\beta$ corresponds to one spatial
dimension. Peschel et al.\cite{peschel-2003-36,PhysRevB.69.075111}
obtain a general result on the reduced density matrix of such a
Hamiltonian
\beq
\rho_s = \mathcal{K} e^{-H'} = \mathcal{K} e^{-\sum_{i,j}H'_{i,j}c_i^\dagger c_j},
\eeq
where $H' = \log(1-C)/C$, $C = (c_i^\dagger c_j)_{i,j}$. Here all the
indices could be vectors when we consider dimension $d > 2$. Their
argument is quite general and applies to not only arbitrary dimensions
but finite temperature as well. And from this $\rho_s$ one can easily
derive the following expression of entanglement entropy:
\beqarr
S & = & \sum_{j = 1}^{L^d}h(\lambda_j), \\
h(x) & = & -\frac{1 + x}{2}\log\frac{1 + x}{2} - \frac{1 -
  x}{2}\log\frac{1 - x}{2},
\eeqarr
which are general and also hold for systems of higher
dimensions. $\lambda_j$'s are eigenvalues of the correlation
matrix $C$. However, a direct computation of $S$ via the
diagonalization of $C$ is highly non-trivial even in the simplest case
as we have seen in previous sections. Wolf's argument is based on the
upper bound and lower bound behaviors of the entropy function
$h(\lambda)$.

His conclusion is that
\beq
c_- L^{d-1} \log L \le S \le c_+ L^{d-1} (\log L)^2,
\eeq
with constants $c_\pm$ depending only on the Fermi sea. This result
requires that the Fermi surface must be regular enough, i.e. not
fractal nor Cantor-like.

Gioev and Klich provide a more specific expression for the scaling
behavior of fermion entanglement entropy by making a connection with
Widom conjecture
\beq
S \sim \frac{L^{d-1}\log L}{(2 \pi)^{d-1}}\frac{1}{12} \int_{\partial
  \Omega} \int_{\partial \Gamma} \abs{n_x \cdot n_p} dS_x dS_p,.
\eeq
This has been verified numerically\cite{barthel-2006-74,li:073103}.

%%%%%%%%%%%%%%%%%%%%%
\subsection{Outlook: Bosonization in Higher Dimensions and Entanglement Entropy}
{\it High Dimension Bosonization}
We have shown basics of one-dimensional bosonization in previous
sections. Bosonization in arbitrary dimensions was
first formulated by Haldane\cite{haldane-1994} (for a review see
\cite{houghton-2000-49}). The basic idea of
bosonization in $d > 1$ dimensions is to
divide the Fermi surface into small segments {\bf S} with height $\lambda$ in
the radial direction and area $\Lambda^{d-1}$ along the Fermi
surface. These two scale must satisfy the following condition:
\beq
k_F \gg \Lambda \gg \lambda.
\eeq
Then we focus on the low energy physics relative to Fermi energy. This
is done by integrating out the high momentum (energy) degrees of
freedom to get the effective Hamiltonian. Working with the effective
Hamiltonian within each segment {\bf S}, we will find ourselves in a
situation similar to that near the Fermi points in one-dimensional
systems. For small momentum transfer $q \ll \lambda$, it is again
possible to pair up the particle-hole excitations in a bosonic way as
in one-dimension case within small correction as long as the
prerequisites are fulfilled.

{\it Possibility of Entanglement Entropy via Bosonization}

The possibility of application of bosonization in higher dimensions
arises from several aspects:

1. According to results in one-dimensional system (given by Korepin
et al.) where correction with respect to the Fermi energy
$\varepsilon_f$ is included, we see that the singular behavior,
i.e. the $\log L$ scaling, only emerges when $\varepsilon_f$ becomes
big enough or the Fermi sea becomes deep enough. This is one of the most
fundamental requirements for the 1-d bosonization to work properly. In
one dimension bosonization becomes exact when we have a infinitely
deep Fermi sea;

2. The agreement on the scaling behavior of entanglement entropy in
free fermion systems and free boson systems;

3. There has been successful application of bosonization in the
calculation of entanglement entropy of two dimensional free Dirac
fields\cite{casini-2005-0507}. Even though in that case the Fermi
surface is absent, it is still nonetheless a strong indication that
bosonization in higher dimensions could work in the presence of Fermi
surface.

The advantage of bosonization approach is that it could take into
account the interactions in arbitrary dimensions.

\bibliographystyle{abbrvnat}

\end{document}